\documentclass[12pt,a4paper,final]{iopart}
\usepackage[breaklinks=true,colorlinks=true,linkcolor=blue,urlcolor=blue,citecolor=blue]{hyperref}
\usepackage{iopams}
\expandafter\let\csname equation*\endcsname\relax
\expandafter\let\csname endequation*\endcsname\relax
\expandafter\let\csname eqalign\endcsname\relax
\expandafter\let\csname fref\endcsname\relax
\expandafter\let\csname Fref\endcsname\relax
\usepackage{slashed} 
\usepackage{caption,subcaption} 
\usepackage{nccmath} 
\usepackage{times}
\usepackage{amsmath} 
\usepackage{amssymb} 
\usepackage{amsopn} 
\usepackage{amsthm} 
\usepackage{verbatim} 
\usepackage{graphicx} 
\usepackage{color} 
\usepackage{booktabs} 
\usepackage{empheq}
\usepackage{relsize}
\usepackage{fancyref}
\usepackage{float}
\usepackage{xcolor,colortbl}
\definecolor{olive}{rgb}{0.3, 0.4, .1}
\definecolor{fore}{RGB}{249,242,215}
\definecolor{back}{RGB}{51,51,51}
\definecolor{title}{RGB}{255,0,90}
\definecolor{dgreen}{rgb}{0.,0.6,0.}
\definecolor{gold}{rgb}{1.,0.84,0.}
\definecolor{JungleGreen}{cmyk}{0.99,0,0.52,0}
\definecolor{BlueGreen}{cmyk}{0.85,0,0.33,0}
\definecolor{RawSienna}{cmyk}{0,0.72,1,0.45}
\definecolor{Magenta}{cmyk}{0,1,0,0}
\definecolor{lcyan}{rgb}{0.6,1,1}
\usepackage[usenames,dvipsnames,pdf]{pstricks}
\usepackage{epsfig}
\usepackage{pst-grad} 
\usepackage{pst-plot} 
\usepackage{pstricks-add}
\usepackage[off]{auto-pst-pdf} 
 
\newcommand{\lp}{\left(} \newcommand{\rp}{\right)} 
  
\newcommand{\ls}{\left[}  \newcommand{\rs}{\right]}

\newcommand{\cd}{\!\cdot\!}
\newcommand{\st}[1]{\slashed{#1}}
\mathchardef\mhy="2D   

\DeclareMathOperator{\J}{J}

\begin{document}
\title[Resonant transitions in strong field particle processes]{Resonant transitions induced in particle processes by the non-perturbative treatment of strong laser fields}
\author{A. Hartin$^1$}
\date{\today}
\ead{anthony.hartin@desy.de}
\address{$^1$CFEL/DESY/University of Hamburg, Luruper Chausee 149, Hamburg 22761, Germany}
\begin{abstract}
Intense field quantum field theory (IFQFT) is used to determine new phenomenological predictions arising from Compton scattering and pair production in a strong laser field. This theory utilizes exact solutions of fermions embedded in external plane waves which carry over to a strong field fermion propagator. When these strong field fermions also interact with probe photons, the transition probability is enhanced in a series of resonances when the kinematics allow the virtual fermion to go on-shell. An analysis of parameters shows that contemporary experiments could already produce these predicted resonances. With appropriate tuning, resonances can be made arbitrarily sharp, leading to precision calculations of resonance locations and widths, and the runing of coupling constants in strong background fields. New tests of QED could be performed leading to ramifications for our understanding of the quantum vacuum. Such tests extend to QFT in background potentials in general, which may throw light on BSM theories such as QFT in curved space-times.
\end{abstract}

\section{Introduction}

Our understanding of the physical vacuum is still relatively limited. In classical physics, the vacuum is merely the unchanging background in which physical phenomena occur. However the quantum vacuum is understood to consist of virtual particles, some of which are charged. An external electromagnetic field can couple to these virtual charges, which in turn affects the behaviour of real particle processes. \\

When a strong electromagnetic field is present, the virtual charges, which form virtual dipoles, start to separate. In the Schwinger limit, where an electric field ($1.3 \times 10^{18}$ V/m) does the work equivalent to separating two rest masses over a Compton wavelength, the vacuum state becomes unstable and the field is predicted to induce vacuum pair production \cite{Dunne09}. \\ 

The Schwinger limit is not expected to be reached with the present generation of planned ultra-intense laser facilities \cite{ELI16,Vulcan06,DiPiaz12,heinzl12}. However, strong fields that reach the Schwinger limit are present in ultra-relativistic heavy ion collisions, where the Coulomb field is altered by 0.1\% on a scale of the Compton wavelength \cite{Rafel98}. Strong background fields are also present in an astrophysical setting near the surface of a magnetar \cite{DauHar83}.\\

There is another way to probe the properties of the quantum vacuum, in the laboratory with today's technology. That is, via the interactions of relativistic electrons, probe photons and strong laser electromagnetic fields. Theoretical models which include the strong laser field non-perturbatively, predict new phenomenology arising from interactions with the polarised vacuum. \\

In a non-perturbative treatment of external fields, standard processes, such as photon emission from electrons, can be interpreted differently. With the external field relegated to the background, the electron can be viewed as becoming unstable and spontaneously radiating. The non-perturbative treatment of an external electromagnetic field, in effect, creates bound electron states and renders the vacuum a dispersive medium. Moreover, when the background field is periodic, the decay/radiation itself shows periodic multi-photon features \cite{Bamber99}. \\

The concept of dispersion in the strong field vacuum, carries over to higher order processes, when a virtual particle is exchanged between initial and final states. The virtual particle probes the strong field vacuum and sees a series of vacuum quasi energy levels. Certain final states arising from given initial states are preferred, resulting in resonant transitions - in analogy to resonant transitions between atomic energy levels \cite{Zeldovich67,Greiner85}. \\

The widths of these predicted resonances correspond to the lifetime of the electron before it decays in the dispersive vacuum/external field background \cite{Oleinik68}. The same phenomena is linked to the strength of the electron self-interaction. This self interaction (or self energy) is altered by the strength of the external field \cite{Ritus72,BecMit76}. \\

The vacuum polarisation is expressed by the photon self-energy. This involves an electron/positron loop which couples to the external field, so vacuum resonances manifest in the vacuum polarisation tensor itself. The measurement of these resonances, via an external field whose direction and strength can be varied, would give us additional information about vacuum polarization effects, complementary to existing schemes \cite{Heinzl06,Valle13}. \\
 
Resonant transitions are predicted also in M{\o}ller scattering \cite{Bos79a}, where a virtual photon is exchanged. Both Compton scattering and two photon pair production in a strong field exhibit resonances through virtue of their virtual particle exchange \cite{Oleinik72,Hartin06}. The universality of the predicted resonant behaviour emphasises that it is a quantum vacuum related effect. \\
  
A successful experimental search of vacuum resonant phenomena would test the general theoretical strategy of non-perturbative quantum field theory in strong background fields. This has implications for approaches to quantum gravity via QFT in curved space-times \cite{Wald10}. This is so, because of the analogy of unstable vacuum states in both theories \cite{FraGitShv91} and for the nonlocality of the particle states for the composite fermion-background field. \\

The divergence structure of IFQFT and its incorporation via regularisation/renormalisation are still open questions with important theoretical work to be done. The treatment of a background external field, modifies the long distance and short distance behaviour of virtual processes. The running of the coupling constant is also modified by coupling to the external field, leading to the non-convergence of the perturbation series at very high field strengths \cite{Narozhnyi79}. \\

A quantum vacuum experiment based on resonant vacuum transitions can be performed in specific laboratories in the coming years after a dedicated theoretical/experimental study is performed. Resonant vacuum transitions can be achieved with parameters far below Schwinger critical levels. Therefore they are accessible for immediate experimental searches. \\

A theoretical study and proposed future experiments are described in this paper. The theoretical model is based on IFQFT and the Furry interaction picture (section \ref{sect:furry}). The strong field processes considered here are the stimulated Compton scattering and stimulated two photon pair production (section \ref{sect:scs}). IFQFT divergences are considered and resonance widths are calculated (section \ref{sect:loops}). Expected experimental signatures for generic experimental parameters within reach today are provided (section \ref{sect:exp}). Natural units with $c=\hbar=1$ are assumed.


\section{Particle processes in strong background fields}\label{sect:furry}

Particle processes taking place in strong fields could be considered in the usual perturbation theory. In that case, the strong field consists of a sea of photons, each of which interacts incoherently with embedded particles. When the external field is provided by an ultra intense laser, the strength of the laser field would then manifest itself in a high number density of laser photons all of which can interact with some primary electron. The probability of multi-photon processes would be enhanced and perturbative calculations would have to proceed to high order. \\

An alternative theoretical framework is to incorporate the strong field as a background at a more fundamental level, so that its influence is analytically complete. The full quantum coupling of the relativistic electron with the external electromagnetic field can be taken into account by solving the Dirac equation in a background potential \cite{BagGit90}. This schema is a semi-non-perturbative quantum field theory (IFQFT). \\


In IFQFT, the Furry interaction picture has the external field, $A^\text{e}$ separated from the quantised gauge field at Lagrangian level. For charged particles, the resulting equations of motion are minimally coupled to the external field (equation \ref{furpic}). The interaction of the field operators of the gauge field and the now "dressed" matter fields, are handled in the usual perturbation theory \cite{JauRoh76}

\begin{align}\label{furpic}
\mathcal{L_{\text{QED}}} &= \bar\psi(i\slashed{\partial}-m)\psi-\frac{1}{4}(F_{\mu\nu})^2-e\bar\psi(\slashed{A}+\slashed{A}^\text{e})\,\psi \notag \\
&\rightarrow\bar\psi^{\text{FP}}(i\slashed{\partial}-e\slashed{A}^\text{e}-m)\psi^{\text{FP}}-\frac{1}{4}(F_{\mu\nu})^2-e\bar\psi^{\text{FP}}\slashed{A}\,\psi^{\text{FP}} \\
&\implies \lp i\slashed{\partial}-e\slashed{A}^\text{e}-m \rp \psi^{\text{FP}}=0 \notag
\end{align}

There are long standing, Volkov solutions $\psi^\text{FP}$, of the minimally coupled Dirac equation, when the external potential can be represented by a plane wave \cite{Volkov35,Hartin11a}. For a real ultra-intense laser experiment, the field can be considered linearly or circularly polarised. Modifications of the exact solutions can incorporate laser pulses as well \cite{SeiKam11}. \\

For an electron of momentum $p_\mu\!=\!(\epsilon,\vec{p})$, mass m and spin r embedded in a plane wave electromagnetic field of potential $A^e_{\text{x}\mu}$ and momentum $k_\mu=(\omega,\vec{k})$, and with normalisation $n_\text{p}$ and Dirac spinor $u_\text{pr}$, the Volkov solution of the minimally coupled Dirac equation is,

\begin{gather}\label{eq:Volkov}
 \Psi^\text{FP}_\text{prx}= n_\text{p}\,E_\text{px}\; u_{\text{pr}}\;e^{- i p\cdot x },\quad n_\text{p}=\sqrt{\mfrac{m}{2\epsilon(2\pi)^3}}\, \\
E_\text{px}\equiv\ls 1 - \mfrac{\slashed{A}^e_\text{x}\st{k}}{2(k\cd p)}\rs e^{-i\mathlarger{\int}^{k\cdot x}\;\mathlarger{\frac{2eA^{e}_\xi\cdot p - e^2A^{e\,2}_\xi}{2k\cdot p}}d\xi} \notag
\end{gather} 

Before these coupled fermion-external field solutions can be used to calculate transition probabilities, they are formed into quantised field operators. First, the solutions are shown to be orthogonal and complete, which they are on both instantaneous time hyperplanes \cite{BocFlo10,Filip85} and on the light-cone \cite{BerVar80a}. Quantisation of the Volkov solutions is then relatively straightforward \cite{BerVar80a,FraGitShv91}. Finally, wave-packets need to be formed. In order to make the assumption that the wave-packet has a sharply peaked momentum, it is best to formulate the wave-packets on the light cone \cite{NevRoh71,IldTor13} which suggests the use of light-cone coordinates \cite{NevRoh71,ChaMa69,KogSop70}. \\

After these preliminary steps, the usual S-matrix or path integral methods can be employed. The LSZ method can be used with Volkov solutions to form the properly interacting theory \cite{LSZ55,IldTor13}. Strong field Feynman diagrams can be drawn and transition probabilities generated. The phenomenology and the techniques used to generate transition probabilities have to be adapted to take into account the new electron-external field states. \\

Assuming all this preliminary theoretical steps are in place, concrete scattering processes involving a fermion propagator to probe the dispersive vacuum, can be considered. 


\section{Stimulated Compton scattering and stimulated two photon pair production}\label{sect:scs}

\begin{figure}[t]
\centering\begin{subfigure}[t]{0.5\textwidth}
\centerline{\includegraphics[width=0.5\textwidth]{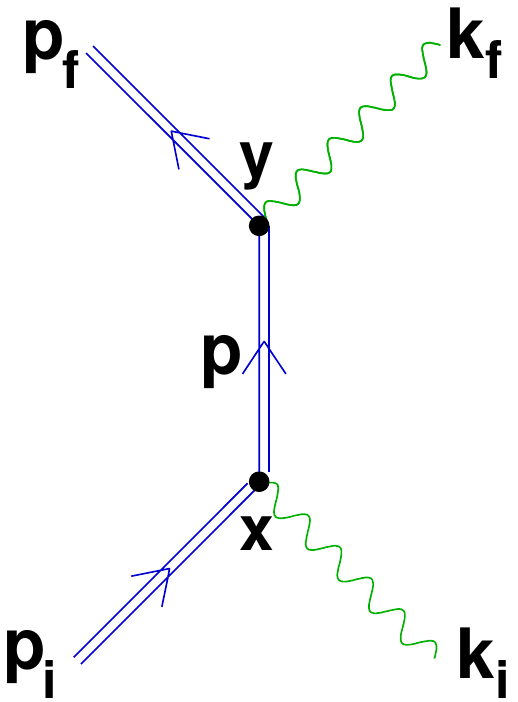}}
\caption{\bf 2nd order IFQFT Compton scattering.}\label{fig:scs}\vspace{0.1cm}
\end{subfigure}\begin{subfigure}[t]{.5\textwidth}
\centerline{\includegraphics[width=0.8\textwidth]{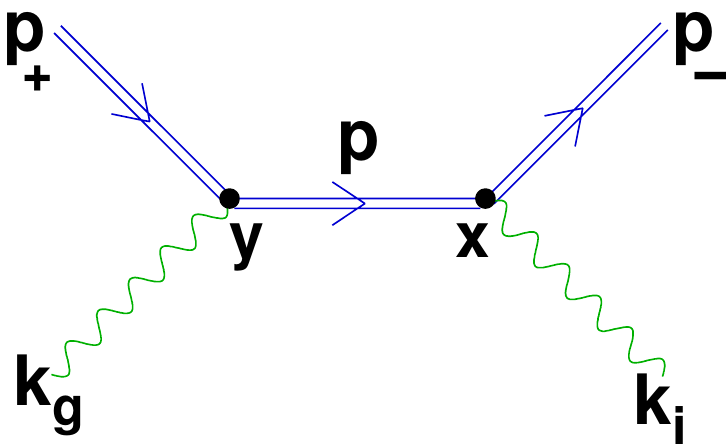}}
\caption{\bf 2nd order IFQFT pair production.}\label{fig:stppp}
\end{subfigure}
\caption{\bf Resonant two vertex processes in the Furry picture. Stimulated Compton scattering (SCS) and stimulated two photon pair production (STPPP) are related through a crossing symmetry.}\label{fig:2ndorder}
\end{figure}

Compton scattering and pair production in a strong background field are to be treated as a strong field Furry picture processes (figure \ref{fig:2ndorder}). Since, as will be seen, the transition probability contains resonances caused by the action of the external field, these processes can be referred to as, stimulated Compton scattering (SCS, figure \ref{fig:scs}) and stimulated two photon pair production (STPPP, figure \ref{fig:stppp}). It is assumed here that the strong field is provided by an ultra-intense laser generating plane waves with a 3-potential $\vec{A}^\text{e}$ and 4-momentum $k$. The initial photons $k_\text{i}$ are provided by a separate probe laser, and a source of high energy photons is needed for the STPPP process. \\

Analysis of the Volkov solution shows that it can be written as a periodic dependence which is amenable to the Floquet theorem and thus form periodic Volkov states. Transitions between these orthogonal, complete Volkov states are favoured, in analogy to transitions between atomic energy levels. Indeed, one can take the analogy further and surmise that there is a physical basis to the Floquet-Volkov states, just as Floquet-Bloch states are based on periodic structures in a solid \cite{Mahmood16}. The Volkov states form in the quantum vacuum, which consists in part, of charged particles which form dipoles to screen real, bare charges. With this picture, the periodic, external electromagnetic field induces structures among virtual dipoles, which in turn lead to Volkov states and the Zeldovich quasi-energy levels \cite{Zeldovich67}. \\

The Fourier transform of the Volkov solution can be taken in order to extract the modes $nk,\; k\equiv(w,\vec{k})$, from the external field potential varying with a period $2\pi L$ \cite{Hartin11a},

\begin{gather}\label{eq:FTVolkov}
 \Psi^\text{FP}_\text{prx}= \sum^\infty_{n=-\infty}\int^{\pi L}_{-\pi L} \mfrac{d\phi}{2\pi L} \; n_\text{p}\,E_\text{px}\; u_{\text{pr}}\;e^{- i (p+lk)\cdot x }\;
\end{gather} 

In a transition from initial to final states, the contribution from the external field is a combination of modes from all contributing Volkov wave functions. Each vertex in a Furry picture Feynman diagram is dressed with a separate sum over contributions. In a multi-vertex process, the complete contribution from the external field will be given by more than one sum.\\

So the transition probability of the two vertex SCS and STPPP processes can be expressed as an overall sum of contributions of photons from the strong field $nk$, with internal contributions $lk$. The action of the external field also induces a momentum/mass shift in fermions $p_\text{i}\rightarrow q_\text{i}$ which is dependent on the relative direction of motion and the intensity of the external field, $a_0$. Thus, the conservation of momentum for the SCS and STPPP processes can be obtained,

\begin{gather} \label{eq:consmom}
q_i+k_i+nk\rightarrow q_f+k_f \quad \text{SCS} \notag\\
k_g+k_i+nk\rightarrow q_{+}+q_{-} \quad \text{STPPP} \\
 n\in\mathbb{Z}\quad m^2=1+a_0^2,\quad q_i=p_i+\mfrac{a_0^2\,m^2}{2k\cd p_i}k,\quad a_0\equiv\mfrac{e|\vec{A}^\text{e}|}{m} \notag
\end{gather}

The condition for resonant transition between quasi-energy levels is established by examination of the process matrix elements, $M^\text{SCS}_\text{f i},\,M^\text{STPPP}_\text{f i}$ consisting of positive and negative energy (electron and positron) Volkov solutions, $\psi^{\text{FP}\pm}$, quantised photons $A_\text{f,i,g}$ and the Furry picture propagator, $G^\text{FP}_\text{yx}$

\begin{gather}
M^\text{SCS}_\text{f i}= \int \text{d}x\, \text{d}y\; \bar \Psi^{\text{FP}+}_\text{fry}\,\bar A_\text{fy}\,G^\text{FP}_\text{yx}\,A_\text{ix}\,\Psi^{\text{FP}+}_\text{isx}\notag\\
M^\text{STPPP}_\text{f i}= \int \text{d}x\, \text{d}y\; \bar \Psi^{\text{FP}-}_\text{fry}\,\bar A_\text{gy}\,G^\text{FP}_\text{yx}\,A_\text{ix}\,\Psi^{\text{FP}+}_\text{isx}\\
\text{where}\quad G^\text{FP}_\text{yx}=\int\mfrac{\text{d}p}{(2\pi)^4}E_\text{py}\;\mfrac{\st{p}+m}{p^2-m^2+i\epsilon}\;\bar E_\text{px}\notag
\end{gather}



The transition probability of these processes yield resonant peaks when the momentum flowing into or out of the virtual fermion matches the energy difference between quasi-energy levels. That is, resonances occur when the propagator $G^\text{FP}$ reaches the mass shell (i.e. when the propagator denominator goes to zero). There are two resonant conditions for each process (only the SCS is written here) corresponding to the direct and exchange channels, and containing internal contributions from the external field $lk$,

\begin{gather}\label{eq:rescond} 
(q_\text{i}+k_\text{i}+lk)^2=m^2(1+a_0^2)\quad  \quad \text{SCS direct channel} \notag\\
(q_\text{i}-k_\text{f}+lk)^2=m^2(1+a_0^2)\quad \text{SCS exchange channel}  
\end{gather}

Without the external field, the mass shell condition can only be obtained for soft photons (the infrared divergence). The IR divergence is partly dealt with by the experimental constraint of finite energy detector resolution (infinitesimally low energy photons can't be detected). In contrast, the SCS and STPPP mass shell conditions can be achieved in multiple circumstances, for experimentally achievable kinematic parameters. With use of the resonance conditions, and with the conservation of energy-momentum, resonance requirements on the initial state photon and the resultant signature of the final states, can be obtained.\\

This resonant feature, of strong field particle processes with a propagator that can go on shell, has been noted before \cite{Oleinik67,Oleinik68,Bos79a,Bos79b,Roshchup96}. In order to perform experiments, it is necessary to do a full calculation of the SCS and STPPP transition probabilities. This is quite a challenging mathematical task due to the relative complexity of the Volkov spinor and phases. However, recent steps have been taken, by use of Fierz transformations of Volkov spinors, that simplify the analytic expressions \cite{Hartin16}. \\

There is further theoretical work to be done. The mass shell condition in the tree level processes have to be turned into resonances by including the line width of the resonant transitions.

\section{Strong field self energy and resonance width}\label{sect:loops}

The transition probability of the tree-level SCS and STPPP processes cannot be calculated at resonance until the propagator poles are dealt with. One possible solution is to calculate the electron self energy and insert it into the propagator denominator via a geometric sum, in analogy to the usual procedure for QED \cite{PesSch95}. \\

There are, however, difficulties with this procedure, in that the electron is coupled not only to its own field, but to the external field as well. The Furry picture self energy does not have a simple dependence on the propagator momentum, and the usual interpretation and absorption of divergent parts is complicated \cite{BecMit76}. \\

\begin{figure}
\centering\begin{subfigure}[t]{0.5\textwidth}
\centerline{\includegraphics[width=0.6\textwidth]{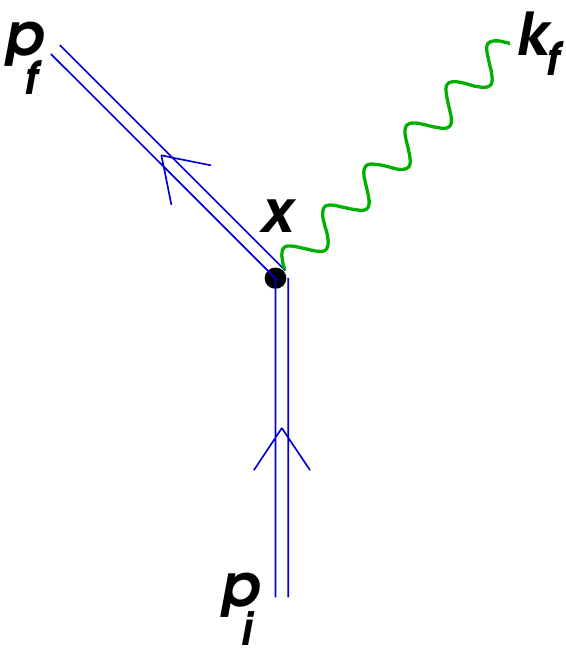}}
\caption{\bf 1st order IFQFT photon radiation (HICS).}\label{fig:hics}\vspace{0.1cm}
\end{subfigure}\begin{subfigure}[t]{.5\textwidth}
\centerline{\includegraphics[width=0.38\textwidth]{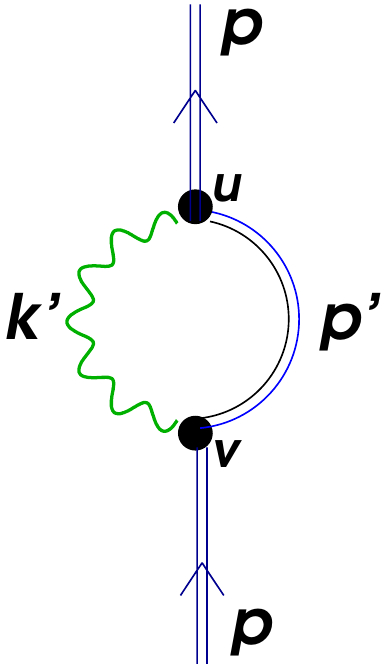}}
\caption{\bf Strong field electron self energy, $\Sigma_\text{p}^\text{FP}$.}\label{fig:ese}
\end{subfigure}
\caption{\bf Furry picture processes related by the optical theorem. The one vertex process is possible through momentum contribution from external field. The self energy couples also to the external field.}
\end{figure}

Another path forward is to calculate a lifetime for the resonant state. The existence, in the Furry picture, of the one vertex photon radiation, in fact suggests a decay of the dressed electron. This interpretation can be physically related to the dispersive nature of the vacuum plus background field, which the Furry picture describes. \\

A vacuum dispersion implies that the intermediate virtual state contains a complex mass shift $M^\text{FP}_\text{p}$ and associated resonance. The resonance will have its center shifted by the real part of the mass shift and the imaginary part will constitute the resonance width. Via the LSZ formula, the coupling constant will also be adjusted from its bare value to its observable value ($e\equiv \sqrt{Z}e_0$) \cite{PesSch95},

\begin{gather}
E_\text{py}\,\mfrac{ie_0^2}{p^2-m_0^2+i\epsilon}\,\bar E_\text{px}\rightarrow E_\text{py}\,\mfrac{iZe_0^2}{p^2-m_0^2+\mathfrak{R}M^\text{FP}_\text{p}+i\mathfrak{I}M^\text{FP}+i\epsilon}\,\bar E_\text{px} \\[4pt]
M^\text{FP}_\text{p}\equiv \Sigma^\text{FP}_\text{p}+... \notag\\[4pt]
\Sigma^\text{FP}_\text{p}=ie^2\sum_\text{spins}\int \text{d}u\,\text{d}v\,\bar E_\text{pu\,}\gamma^\mu\, G^\text{FP}_\text{uv} \,D^\text{FP}_\text{vu} \,\gamma_\mu \, E_\text{pv} \notag
\end{gather}

The LSZ formula relates the one particle irreducible (1PI) diagrams to the forward scattering. The largest leading term in the 1PI sum, the Furry picture self energy, $\Sigma^\text{FP}_\text{p}$ (figure \ref{fig:ese}) is related to the HICS radiation process (figure \ref{fig:hics}). To leading order then, the resonance width $\Gamma$ is,

\begin{gather}
\mathfrak{I}M^\text{FP}_\text{p}|_\text{leading order} =\mathfrak{I}\Sigma^\text{FP}_\text{p}=2 W^\text{HICS}\equiv\Gamma
\end{gather}

The imaginary and real parts of the mass correction $M^\text{FP}_\text{p}$ are related to each other through a dispersion relation, and both are related to the observable coupling constant. Therefore, a precise experimental determination of Furry picture, resonance poles and widths, would provide sensitive new tests, both for QED and for the more general quantum field theories in background potentials. \\

The procedure of dealing with self energies and including them in tree level processes is well understood in normal perturbation theory. However, there are still open questions regarding the procedure in the Furry picture.\\

The first question regards the renormalisation, which is necessary due to theory being formulated for masses and charges of non-interacting fields. This remains necessary even if the interaction with the external field is exact. The procedure requires regularisation and absorption of UV divergences into observable mass and charge. \\

The second question regards the UV divergence itself. Physically, it is understood as arising from BSM physics, that modifies the theory at small distance scales. In the Furry picture, with a background field which modifies the interactions at all scales, it is possible that the UV divergences are automatically regularised. An investigation is underway.\\

The third question regards the running of the coupling constant as the external field strength grows towards the Schwinger critical value. Since the external field couples with the self energy to the electron, the field strength plays a part in the coupling constant. At a certain level, the sum of 1PI diagrams appears to no longer converge \cite{Narozhnyi79}. \\

Another theoretical challenge is the instability of the vacuum state as the spontaneous pair production threshold is reached. This, combined with the nonlocality of the Volkov solution in both momentum and position space, has direct analogy to similar challenges in QFT in curved spacetimes \cite{Wald10}. \\

These open theoretical questions are currently under consideration. In any case, one can determine experimentally where in parameter space resonant transitions are likely to occur. Experimental designs and requirements can already be explored with the aim of shedding light on theoretical predictions and challenges. 

\section{Experimental schema for resonance detection}\label{sect:exp}

Intense laser/electron beam interactions have already been performed successfully, in order to produce strong field signatures of first order Furry picture processes. The strong field radiation process (figure \ref{fig:hics}) was induced in the head-on collision of 46.6 GeV electrons and an optical laser focussed to an intensity of order $10^{18} \text{ Wcm}^{-2}$ \cite{Bamber99}. \\

\begin{figure}
\centering\begin{subfigure}[t]{0.5\textwidth}
\centerline{\includegraphics[width=0.7\textwidth]{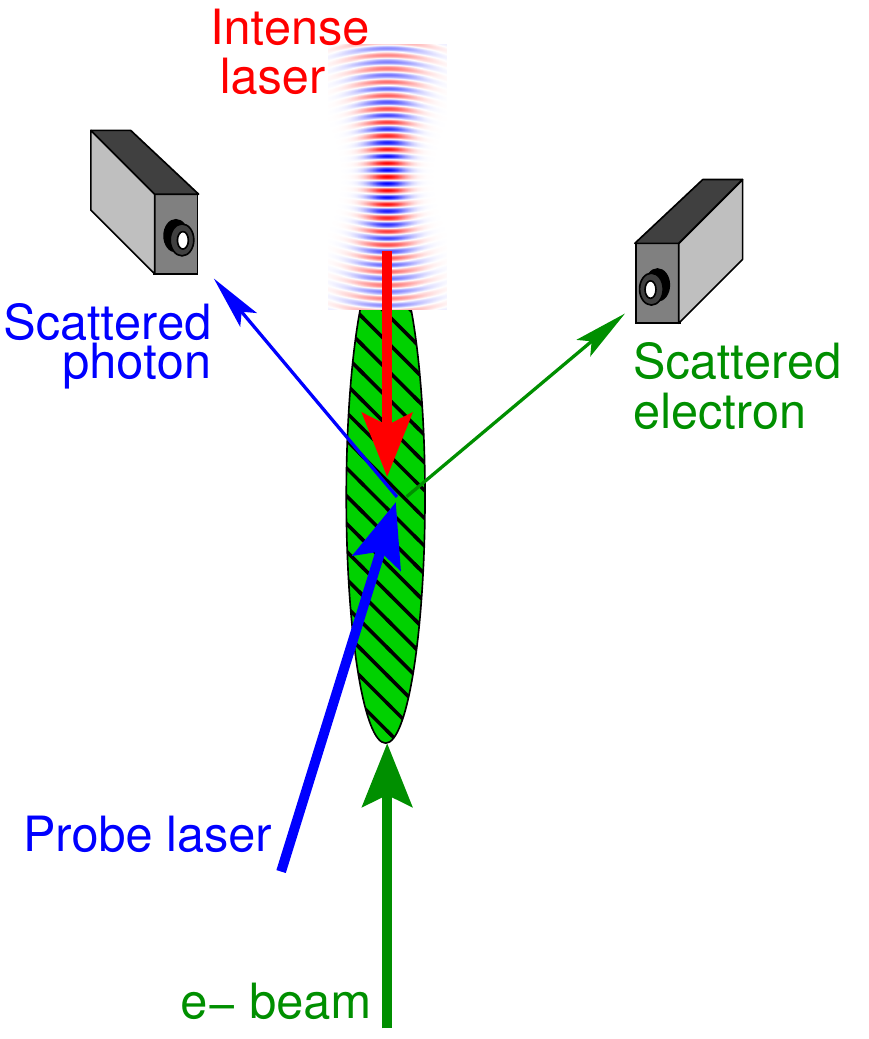}}
\caption{\bf Experimental set up for the SCS process.}\label{fig:scsexptconfig}\vspace{0.1cm}
\end{subfigure}\begin{subfigure}[t]{.5\textwidth}
\centerline{\includegraphics[width=0.9\textwidth]{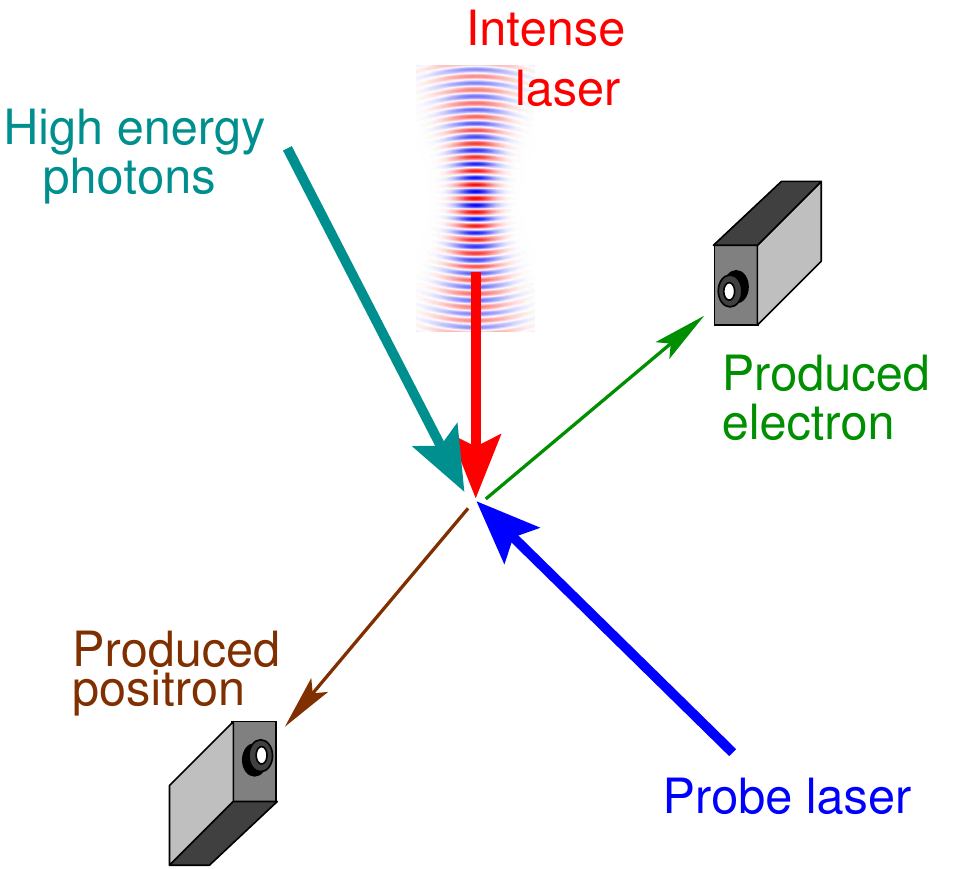}}
\caption{\bf Experimental set up for the STPPP process.}\label{fig:stpppexptconfig}
\end{subfigure}
\caption{\bf Experimental schematic for the production of vacuum resonances in second order Furry picture processes.}
\label{exptconfig}\end{figure}

In order to experimentally produce the resonances of the SCS process, a similar, near head-on collision can be set up, preferably with primary inverse Compton scattered photons directed away from the final state region of interest. Resonances can then be scanned over by introducing a probe laser whose direction and/or energy can be varied. Scattered particles are captured with suitable detectors which can scan the range of final state angles and energies (figure \ref{fig:scsexptconfig}). \\

For the STPPP process, a source of high energy photons sufficient to reach the threshold of pair production, should interact with the strong laser at some small incident angle to inhibit one photon pair production processes. Scans can once again be made over probe photon energies and angles to locate resonant STPPP pair production (figure \ref{fig:stpppexptconfig}). \\

The analysis continues here only for the SCS process. Similar relations and features will pertain to the STPPP process (given that production threshold is reached) since STPPP is related to SCS through a crossing symmetry \cite{Hartin06}. \\

In terms of experimentally available parameters, one can assume that a table top, optical laser focussed to $10^{19} \text{ Wcm}^{-2}$ is routine. This corresponds to a strong field intensity parameter of $a_0\gtrsim 1$, though intensity can be tuned lower with longer pulse lengths, or less stringent focussing. \\

\begin{figure}
\centering
\includegraphics[width=0.3\textwidth]{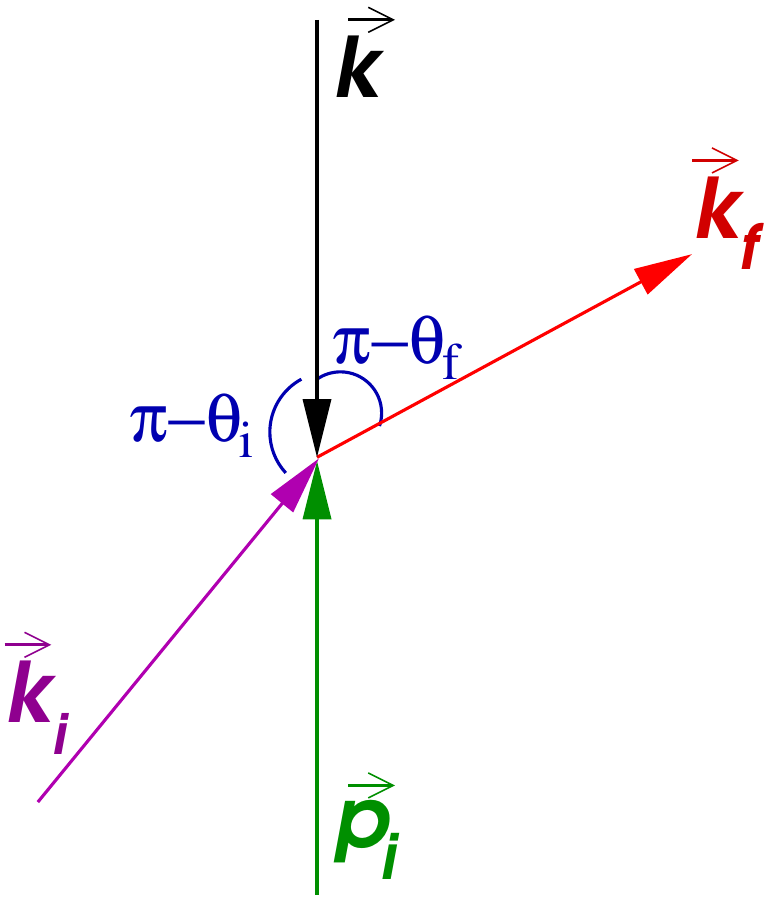}\caption{\bf probe and radiated photon angles in the SCS process.}
\label{fig:scsang}\end{figure}

For the SCS process, relativistic electron beams are also necessary. Electron energy is assumed to vary over a large range, with resultant radiated photons generally compressed in a forward $1/\gamma$ cone. The probe laser should be a tunable optical laser, with a corresponding energy $\omega_\text{i}$ and varying angle of incidence to the external field propagation direction, $\theta_\text{i}$ (figure \ref{fig:scsang}). There is now enough information to determine conditions for resonance. \\

For relativistic electrons ($\gamma,\beta$) colliding head-on with a strong laser of intensity $a_0$ and energy $\omega$, the condition for the nth level resonance can be obtained. The direct channel gives a resonant probe laser incident angle, and the exchange channel gives a resonant radiated photon angle,

\begin{figure}
\centering
\includegraphics[width=0.8\textwidth]{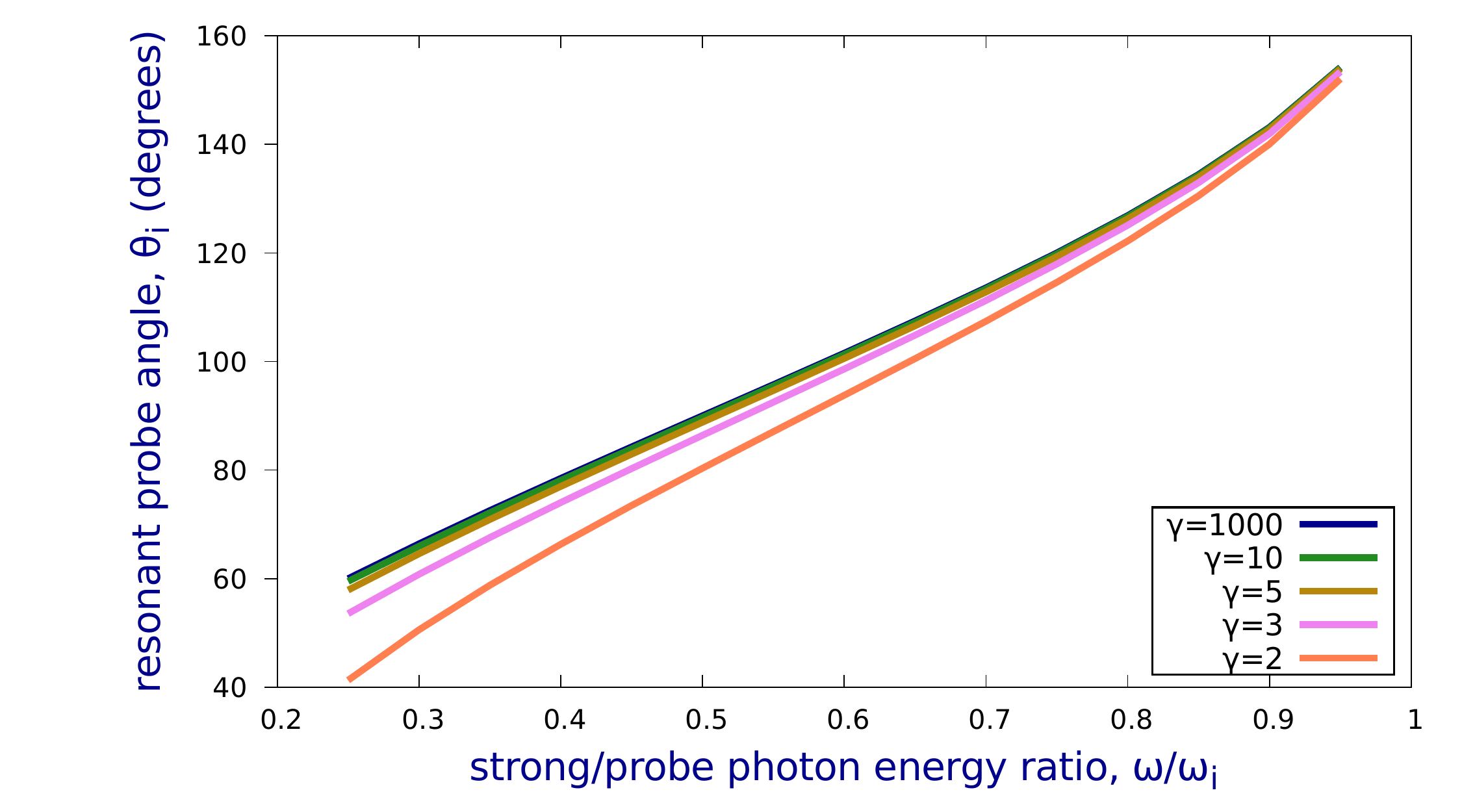}\caption{\bf Resonant probe angle vs photon energy ratio for relativstic $\gamma$ electrons.}
\label{fig:theti}\end{figure}

\begin{gather}\label{fig:resangles}
\cos\theta_i\approx\mfrac{1+\beta +(1+\beta)^2l\omega/\omega_i +a_0^2/2\gamma^2}{\beta(1+\beta)-a_0^2/2\gamma^2} \quad \text{direct channel}\notag\\
\cos\theta_f\approx\mfrac{1+\beta -(1+\beta)^2l\omega/\omega_f +a_0^2/2\gamma^2}{\beta(1+\beta)-a_0^2/2\gamma^2} \quad \text{exchange channel}
\end{gather}

\begin{figure}
\centering
\includegraphics[width=0.8\textwidth]{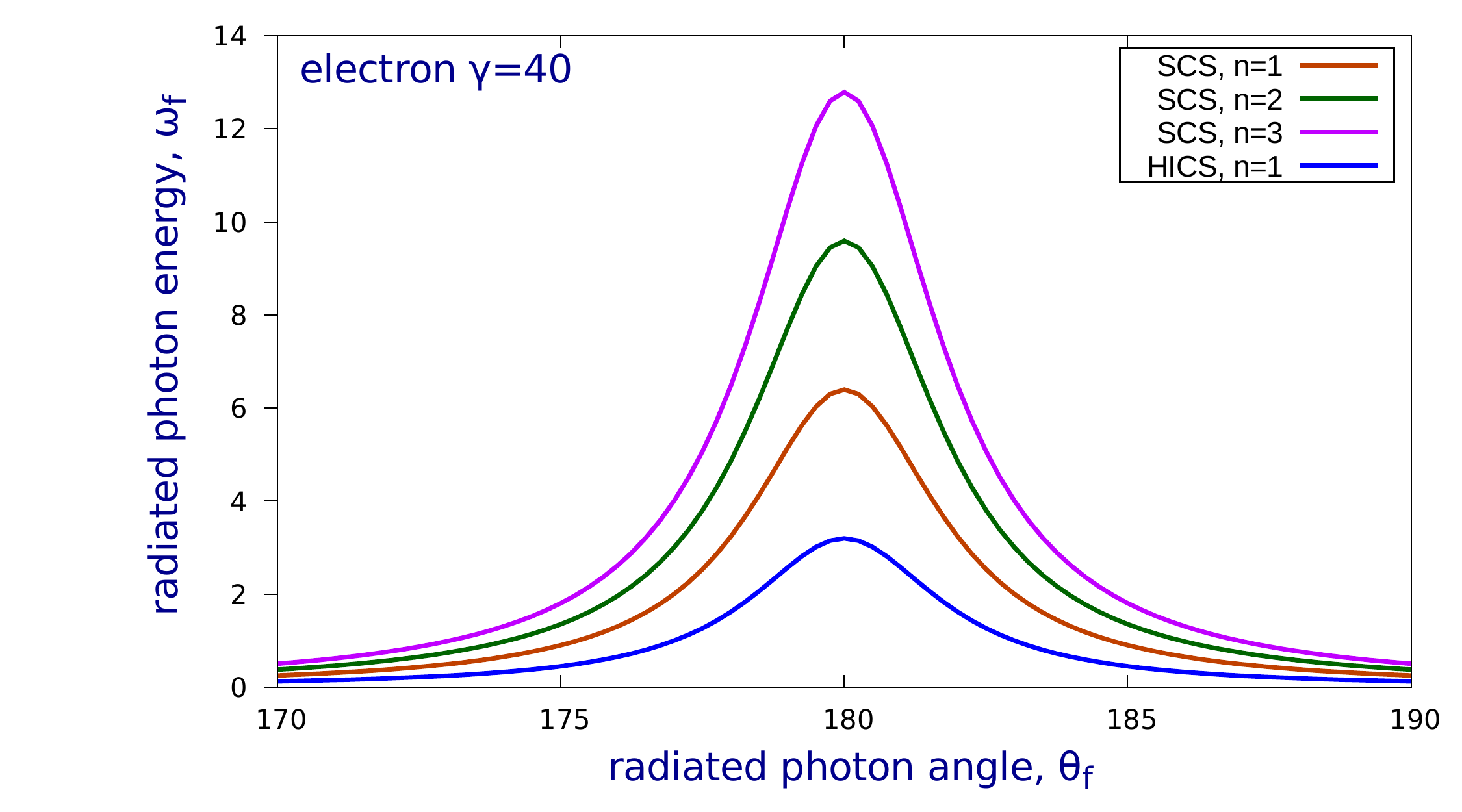}\caption{\bf Radiated photon energy vs radiation angle at resonance.}
\label{fig:thetf}\end{figure}

The direct channel resonance condition depends only on initial state parameters. Assuming $a_0\approx 1$ for the strong laser, and a broad range of electron energies, the tuning of the initial probe photon energy results in a clear resonant incident angle (figure \ref{fig:theti}). The signature of the radiated photons at the direct channel resonance is given by the conservation of energy-momentum and is a smooth distribution corresponding to a forward cone relative to the electron direction. The energy of the radiated photon steps up for each external field mode corresponding to the overall contribution from the external field $nk$ (figure \ref{fig:thetf}). \\

\begin{gather}
\omega_f=\mfrac{(n+1)\,\omega\,\gamma(1+\beta)}{\gamma(1\!+\!\beta\cos\theta_\text{f})\!+\!\ls n\omega\!+\!\frac{a_0^2}{\gamma(1+\beta)}\rs(1\mhy\cos\theta_\text{f})+\omega_\text{i}\!\ls1\mhy\cos(\theta_\text{i}\!+\!\theta_\text{f})\rs}
\end{gather}

The exchange channel resonance, by contrast, will show resonant radiated photon angles. The overall resonance structure is a combination of direct and exchange channels in the complete SCS transition probability. The width and height of the resonances, whether they are scanned over angles or probe energy, is dependent on both the numerator of the transition probability and the imaginary part of the self energy correction appearing in the denominator \cite{Hartin06}. \\

An estimate of the transition probability at resonance can be made by comparing the tree level propagator denominator to the resonance width. The resonance width is determined from the imaginary part of the self energy. For an strong laser intensity $a_0\approx 1$, the resonance width can be given approximately in terms of the QED coupling constant $\alpha$, and the scalar product of field momentum and virtual electron momentum $\rho\equiv 2k\cd p$ \cite{BecMit76},

\begin{gather}
\Gamma\approx 0.29\,\alpha\,\rho\,a_0^{1.86}
\end{gather}

\begin{figure}
\centering
\includegraphics[width=0.8\textwidth]{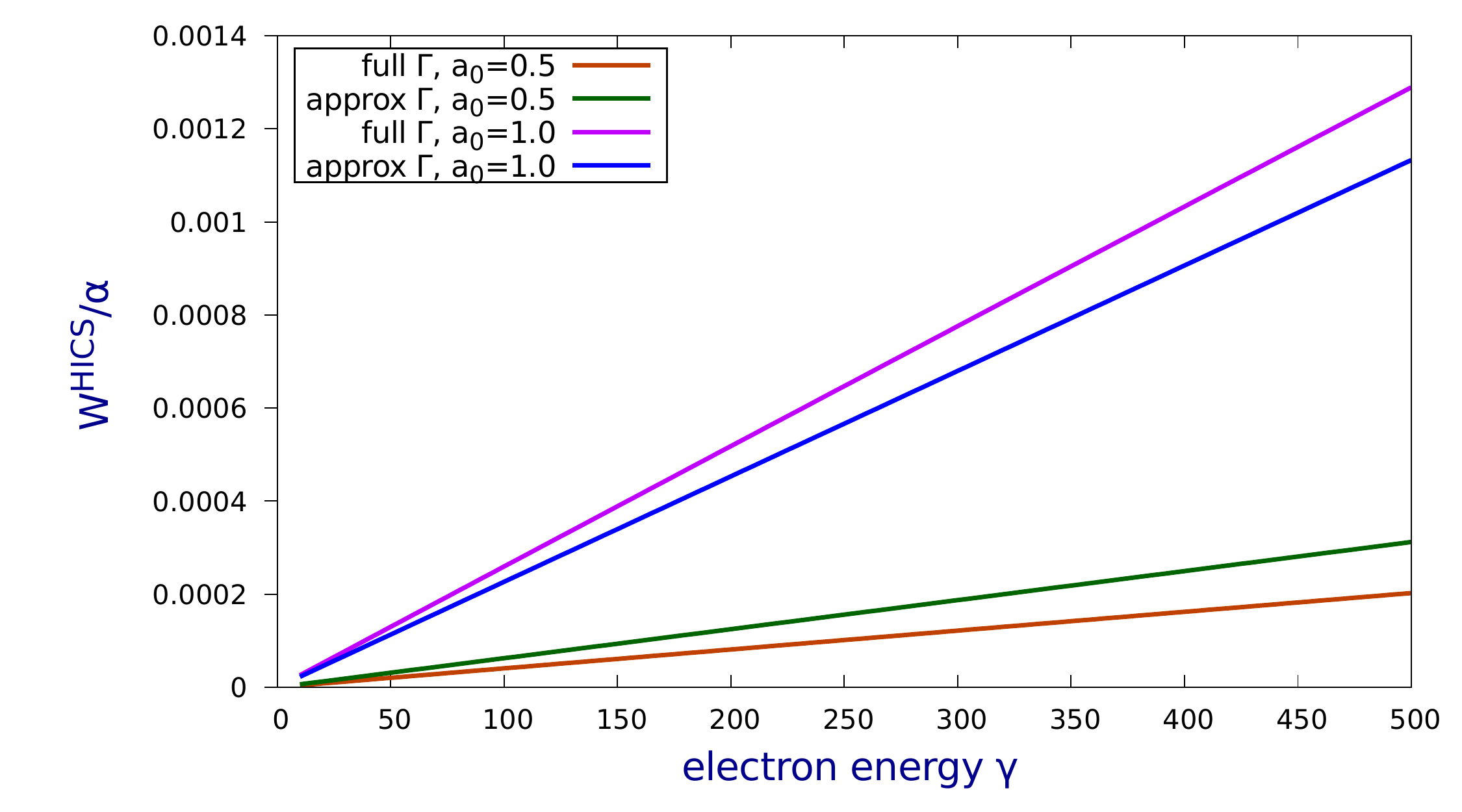}\caption{\bf Comparison of exact and approximate expressions for the resonance width, $\Gamma\equiv \mathfrak{I}\Sigma^\text{FP}_\text{p}\equiv W^\text{HICS}$.}
\label{fig:esecomp}\end{figure}

This approximation assumes a circularly polarised external field, and arises partly from the Fourier transform of the Volkov phase which leads to the appearance of Bessel functions. The approximate expression for $\Gamma$ can be compared to the full expression obtained from the equivalence between the imaginary part of the self energy and the HICS transition rate\label{eq:whics}, and the agreement is reasonable for an order of magnitude estimation (figure \ref{fig:esecomp}).

\begin{gather}\label{eq:whics}
W^\text{HICS}=\mfrac{\alpha}{2} \sum_{n=1}^\infty\int^{u_n}_0 \!\mfrac{du}{(1+u)^2}\ls \mfrac{4}{a_0^2}\J_n^2 - \mfrac{2+2u+u^2}{(1+u)}\lp\J_{n-1}^2+\J_{n+1}^2-2\J_{n}^2 \rp \rs \\
\text{with Bessel function arguments, }z=\mfrac{2\,n\,a_0}{\sqrt{1+a_0^2}}\sqrt{\mfrac{u}{u_n}\lp 1-\mfrac{u}{u_n}\rp} \quad\text{and } u_n=\mfrac{n\rho/m^2}{1+a_0^2}\notag
\end{gather}

To determine the transition probability at resonance, the resonance width $\Gamma$, must be compared with the tree level propagator denominator, which itself is roughly of the order $\rho$. Taking $\rho$ out as a common factor, the propagator denominator is written as a term of order one, and the contribution from the resonance width,

\begin{gather}
\rho\ls \mfrac{q^2-m_\ast^2}{\rho}+\text{i}\, 0.29\,\alpha\,a_0^{1.86}\rs
\end{gather}

Thus at resonance, the SCS transition probability (with $a_0\approx 1$) will increase by a factor of $(1/0.29\alpha)^2$. Given that the two vertex SCS process is a factor of $\alpha$ smaller than the one vertex HICS process, at resonance $W^\text{SCS}$ will exceed $W^\text{HICS}$ by about three orders of magnitude. As the intensity $a_0$ exceeds one, the effect becomes larger. \\

As the resonance grows in height, it of course narrows, which is a factor limiting the number of radiated photons. The estimation is also mitigated by the phase factors and the numerical value of the SCS trace at resonance. A full calculation provides the precise numbers \cite{Hartin06}. \\

\begin{figure}[htb]
\centering
\includegraphics[width=0.8\textwidth]{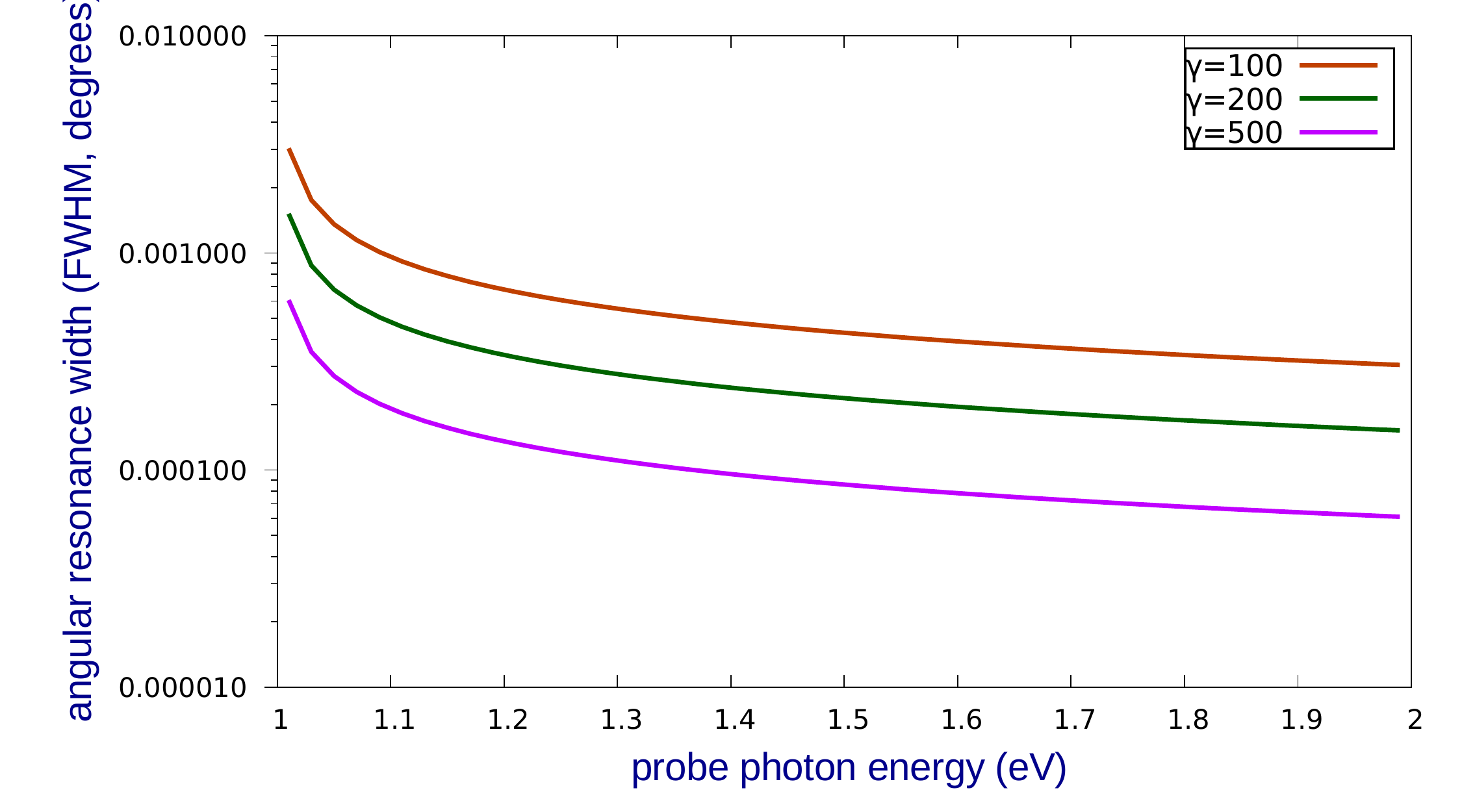}\caption{\bf Direct channel angular resonance width for 1eV laser photons}
\label{fig:fwhm}\end{figure}

The resonance width yields the angular resolution of the differential cross-section resonance using the expressions of equation \ref{fig:resangles}. Using the condition for the direct channel resonance, the peak angular resolution (FWHM) is up to 0.003 degrees for a 200 MeV electron, when the ratio between probe photon energy and laser photon energy approaches unity (figure \ref{fig:fwhm}). The resonance broadens as the electron energy decreases. For higher electron energies, the angular resolution decreases towards the limit of what was experimentally resolvable in similar past experiments \cite{McDonald91}. Correspondingly, a small angular resolution will give a precise measurement of the resonance location. \\

In a real experiment, there will be other factors that will play a role. Specifically, the emittance of the electron beam and the fact that the strong laser field will be provided by a pulse, are factors that will broaden SCS resonances. Other factors will be the resolution of detectors and background processes. Nevertheless, the magnitudes are such that a dedicated experiment should be able to distinguish these new predicted resonant effects. 

\section{Conclusion}\label{sect:concl}

Quantum field theory with the non-perturbative inclusion of strong electromagnetic fields, predicts interesting new phenomena. The Schwinger critical field, which is not yet in reach, envisages the creation of electron-positron pairs from the vacuum. For background fields of more moderate intensity, that can be produced today, higher order processes with a charged virtual particle reach the mass shell and should display large resonant transitions for tuned experimental parameters. \\

These "vacuum resonances" can be understood in analogy to resonant atomic transitions. A quasi-energy level structure is set up in the vacuum by the strong electromagnetic field. The physical basis for these quasi-energy levels are likely to be the virtual charges themselves which respond to applied fields and preference certain virtual particle propagation states. \\

The existence of virtual charges is already known through well established phenomena like the Lamb shift. The Lamb shift, however is an atomic phenomena. The second orderd IFQFT processes outlined here, would manipulate vacuum charges using a circularly or linearly polarised electromagnetic field provided by a strong laser. Given a careful selection of parameters, the experimental signals should be clear and their discovery would constitute a new test of QED. \\

Indeed, the experimental test outlined in this paper would be a test not just of QED, but of intense field quantum field theories, in general.  This class of theories predict a range of new phenomena including a shift in mass of particles which couple to the strong field. These theories predict also that the running of the coupling constant is modified. All of these predictions are amenable to experimental investigation and the current scope for new discoveries is large, given a dedicated search. \\

There are also implications for a theory of quantum gravity which formulates QFT in a curved background space. This theory is closely related to strong field QFT and any experimental confirmation of strong field QFT predictions, particular as regards its virtual particle or vacuum state, will be highly informative. \\

A scenario for experimentally detecting SCS resonances with experimentally available parameters was briefly sketched. The analysis carries over for the STPPP process, indeed for any IFQFT process which couples to the external field and involves virtual particle exchange. The location of resonances in parameter space is easily established with the momentum flowing into or out of the exchanged virtual particle. For tuned parameters, the resonances will appear as sharp peaks in scans over probe photon energy and angle of incidence. \\

A full transition probability calculation uses Volkov solutions of the electron embedded in the external strong electromagnetic field. Recent work has simplified the procedure for obtaining the complete analytic expression \cite{Hartin16}. Resonance line widths are determined theoretically through the non-perturbative 1PI processes, and included via the LSZ formula. \\

Additional modelling is required for conditions in the laboratory. Real, ultra-intense lasers are likely to be linearly polarised and pulsed. The strong field calculations can easily be adjusted to take this into account. Additional work to consistently apply renormalisation and determine the nature of UV divergences in IFQFT is under way. This will enable the consistent inclusion of loops and will predict resonance widths and shifts from their tree-level parameters. \\

In this current period of uncertainty about which direction nature goes beyond the standard model, the predictions and experiments outlined here promise to give new insights into our current models and perhaps indicate a way forward.

\section*{Acknowledgments}
The author acknowledges funding from the Partnership of DESY and Hamburg University (PIER) for the seed project, PIF-2016-53.

\section*{References}
\bibliographystyle{unsrt}
\bibliography{/home/hartin/Physics_Research/mypapers/hartin_bibliography}


\end{document}